\documentclass{article}
\usepackage{emulateapj}
\usepackage{amsmath}
\usepackage{amssym}
\usepackage{psfig}
\usepackage{times}
 
\input{epsf}
 
\def\sT{\sigma_{\rm T}}
\def\tT{\tau_{\rm T}}
\def\beq{\begin{equation}}
\def\eeq{\end{equation}}
\def\dd{{\rm d}}
\def\epL{\varepsilon_{\rm L}}
\def\ep{\varepsilon}
\def\Tbb{T_{\rm bb}}

\newbox\grsign \setbox\grsign=\hbox{$>$} \newdimen\grdimen \grdimen=\ht\grsign
\newbox\simlessbox \newbox\simgreatbox \newbox\simpropbox
\setbox\simgreatbox=\hbox{\raise.5ex\hbox{$>$}\llap
     {\lower.5ex\hbox{$\sim$}}}\ht1=\grdimen\dp1=0pt
\setbox\simlessbox=\hbox{\raise.5ex\hbox{$<$}\llap
     {\lower.5ex\hbox{$\sim$}}}\ht2=\grdimen\dp2=0pt
\setbox\simpropbox=\hbox{\raise.5ex\hbox{$\propto$}\llap
     {\lower.5ex\hbox{$\sim$}}}\ht2=\grdimen\dp2=0pt
\def\simgt{\mathrel{\copy\simgreatbox}}
\def\simlt{\mathrel{\copy\simlessbox}}

\lefthead{BELOBORODOV \& POUTANEN}
\righthead{POLARIZATION NEAR THE LYMAN EDGE IN QUASARS}
 
\begin{document}

\title{On the Origin of Polarization near the Lyman Edge in Quasars}
 
\author{Andrei M. Beloborodov\altaffilmark{1} and Juri Poutanen}
\affil{Stockholm Observatory, Saltsj\"obaden, S-133 36, Sweden}

\altaffiltext{1}{Also at Astro-Space Center of Lebedev Physical Institute,
Profsoyuznaya 84/32, 117810 Moscow, Russia}

\begin{abstract}
Optical/UV radiation from accretion disks in quasars is likely to be partly 
scattered by a hot plasma enveloping the disk.
We investigate whether the scattering may produce the steep rises in 
polarization observed blueward of the Lyman limit in some quasars. 
We suggest and assess two models. In the first model, primary disk 
radiation with a Lyman edge in absorption passes through a static ionized 
``skin'' covering the disk, which has a temperature $kT\sim 3$ keV and 
a Thomson optical depth $\tT\sim 1$. Electron scattering in the skin smears 
out the edge and produces a steep rise in polarization at $\lambda<912$ {\AA}. 
In the second model, the scattering occurs in a hot coronal plasma outflowing 
from the disk with a mildly relativistic velocity. We find that the second 
model better explains the data. The ability of the models to fit the observed 
rises in polarization is illustrated with the quasar PG~1630+377.
\end{abstract}



\section{Introduction}

Steep rises in polarization blueward of the Lyman limit have been detected 
in 3 of 10 observed quasars (Impey et al. 1995; Koratkar et al. 1995).
Especially challenging is the case of PG 1630+377 where the polarization 
increases from $\sim 2$\% at 1000 {\AA} to $\sim 20$\% at 700 {\AA}.
Attempts to explain such a steep rise have not been successful (e.g., Blaes 
\& Agol 1996; see Koratkar \& Blaes 1999 for a review). One toy model fitting 
the data was suggested by Shields, Wobus, \& Husfeld (1998). They assume, 
however, an arbitrary large {\it ad hoc} jump in polarization at $\lambda=912$ 
{\AA} in the disk comoving frame and then fit the observed rise by the 
relativistically smeared-out jump.

A possible physical mechanism producing the rises in polarization is 
Compton upscattering of the disk radiation by a hot plasma.
Assuming that the disk emits UV radiation with a Lyman edge in absorption, 
one finds that the upscattering (1) smears out the Lyman edge and 
(2) polarizes the radiation. The upscattered radiation 
dominates the spectrum blueward of the edge, and this leads to the rise in 
polarization at $\lambda<912$ {\AA}. 

We investigate here two scenarios:
\medskip

\noindent
A. {\it Multiple scattering in a static slab with Thomson optical depth 
$\tT\sim 1$.} The scattered radiation is limb-brightened in the slab and the 
multiply scattered radiation acquires a polarization parallel to the disk 
normal (Sunyaev \& Titarchuk 1985). 
\medskip

\noindent
B. {\it Scattering in a mildly relativistic outflow of optical depth $\tT<1$.} 
Relativistic aberration causes a limb-brightening of the disk radiation in
the outflow rest frame. This effect leads to a parallel polarization, 
$p\sim 20$\%, of the scattered radiation (Beloborodov 1998, hereafter B98).
\medskip

Transfer of polarized radiation in a static slab corona of a high temperature,
$kT\sim 100$ keV, was previously studied in detail (e.g., Haardt \& Matt 1993; 
Poutanen \& Svensson 1996; Hsu \& Blaes 1998).
It has been shown that the corona produces parallel polarization rising 
toward short wavelengths. Possible relevance to the UV polarization 
rises in QSOs has been assessed by Hsu \& Blaes (1998) who concluded that the 
predicted rise is not steep enough to explain the data. 

In this Letter, we argue that the scattering in a layer at a Compton
temperature, $kT_{\rm C}\sim 3$ keV, would generate a much steeper rise in
polarization as compared to a hot static corona with $kT\sim 100$ keV. 
The layer may be associated with an ionized ``skin'' of an accretion disk
heated by the corona emission.
In \S 2, we perform calculations of radiative transfer in the skin
demonstrating the production of steep rises in polarization.

Then, in \S 3, we investigate the outflow model. Previously, it was suggested 
as a possible mechanism producing parallel optical polarization (B98). 
We now find that the model can reproduce the rises in UV polarization.
The results are summarized in \S 4.


\section{Comptonization in a static slab}

\subsection{Production of parallel polarization}

Let semi-isotropic radiation propagate into an ionized slab from an underlying 
disk. With increasing height, the attenuated intensity of unscattered 
radiation, $I_0$ (the 0-th scattering order), gets suppressed at large angles 
with respect to the normal because the slab has large optical depths at these 
angles. The limb-darkening of $I_0$ 
results in perpendicular polarization in the first scattering order. 
In the second order, photons come from the slab itself, and their intensity, 
$I_1$, is limb-brightened if the slab optical depth $\tT\simlt 1$. 
The radiation scattered twice, $I_2$, then acquires a parallel polarization. 
The limb-brightening progresses with successive scatterings and then saturates
as the intensity $I_N$ approaches the eigen-function of the transfer problem 
(Sunyaev \& Titarchuk 1985). The resulting parallel polarization progressively 
increases with scattering orders and then saturates. 
For $\tT\sim 1$, the saturation occurs after $\sim 5$ scatterings. 

In a hot slab, the scattering is accompanied by a photon blueshift. 
After $N$ scatterings, a photon of initial energy $\ep$ acquires an energy of 
$\sim(1+4kT/m_ec^2)^N\ep$ (see, e.g., Rybicki \& Lightman 1979). One thus 
expects a strong impact of the slab on both the disk spectrum and the 
polarization. 
If the original UV spectrum has a Lyman edge in absorption then the edge 
energy, $\epL=13.6$ eV, delineates two regions in the emerging spectrum. 
At $\ep>\epL$, the upscattered radiation dominates the observed flux, 
and the further away from the edge, the higher are the scattering orders 
contributing to the spectrum. The radiation is therefore parallelly polarized 
at $\ep\gg\epL$. By contrast, at $\ep<\epL$, all scattering orders contribute 
to the spectrum, and this leads to a small net polarization.

A slab with temperature $kT\sim 100$ keV produces a gradual growth in parallel 
polarization toward high energies, so that $p$ reaches $\sim 10-15$\% at 
$\ep\simgt 1$ keV (Hsu \& Blaes 1998). The polarization maximum is blueshifted 
far away from the edge because the gain in energy per scattering is large,
$\Delta\ep/\ep\sim 4kT/m_ec^2\sim 4/5$. The high $T$ is thus the reason for 
the slow growth of $p$ with respect to wavelength.
If the Comptonization occurred in a slab of a relatively low temperature, then 
a large number of scatterings would be accompanied by a modest blueshift, and
$p(\lambda)$ at $\lambda<912$ {\AA} would be much steeper.
With this motivation, we consider a transition layer (``skin'') 
between the UV disk and the corona as a possible location of the Comptonization.
The layer is Compton heated by the corona emission to an equilibrium 
temperature $kT_{\rm C}$ of a few keV and has an optical depth $\tT\sim 1$
(e.g., \.Zycki et al. 1994).
The Kompaneets' $y$-parameter of the layer is $4(kT/m_ec^2)\tau_T^2\ll 1$, 
which implies a very soft spectrum of upscattered photons.

If the disk has a patchy hot corona then the bulk of UV radiation comes 
directly from the disk without being scattered in the corona
(Haardt, Maraschi, \& Ghisellini 1994). At the same time,
the whole UV disk may be covered by the ionized skin. Then the skin may 
dominate the Comptonized spectrum at modest energies, while the patchy 
corona dominates the hard X-ray spectrum. The skin may contribute 
to the soft excess observed in quasar spectra.

We model the skin as a slab of optical depth $\tT\sim 1$ and temperature 
$kT\sim 1-10$ keV. Below the slab, we assume a blackbody source with 
temperature $k\Tbb=3$ eV. The Lyman edge in the source spectrum is modeled as 
a sharp jump down to zero intensity at 912 {\AA}.
In the calculations of radiative transfer in the slab, we use the iterative 
scattering method described in detail in Poutanen \& Svensson (1996).

\subsection{Results}

The results for $\tT=1$ and $kT=$1, 3, and 10 keV are presented in Figure 1 
for a disk inclination $\cos\theta=0.23$. The polarization is close to its 
maximum at this inclination.
With increasing $\tT$, $p$ decreases, changing its sign, and in the limit 
$\tT\gg 1$ one arrives at the standard Chandrasekhar-Sobolev perpendicular 
polarization. An increase in $T$ leads to a larger energy gain per 
scattering that makes the polarization rise less steep. 
On the other hand, a decrease in $T$ and/or $\tT$ leads to 
a reduction of the flux blueward of the Lyman edge. 
This is a general tendency of a Comptonization model: dramatic rises in 
polarization imply modest fluxes blueward of the edge.  
A compromise is achieved when $\tT\sim 1$ and $kT\sim 3$ keV. In this case,
the predicted behavior of the flux and polarization resembles the data on 
PG 1630+377 (see Fig. 1). The edge is smeared-out at large inclinations,
while it is strongly pronounced at modest inclinations (see Fig. 2). Additional 
smearing-out may be present due to relativistic effects near a Kerr black hole.

A robust feature of the model is that $p$ changes sign across the edge, i.e.,  
the small polarization redward of the edge is perpendicular to the disk normal 
(see also Haardt \& Matt 1993).
By contrast, the data favor the same position angle blueward and redward of 
the edge (see Koratkar et al.  1995). The constancy of the position angle 
is better seen in PG 1222+228 (Impey et al. 1995).
Besides, optical polarization measured in many objects tends to be 
parallel to the disk normal (e.g., Antonucci 1992). The predicted perpendicular
redward polarization thus appears to contradict the data.

\medskip

\smallskip
\centerline{
\epsfxsize=8.4cm {\epsfbox{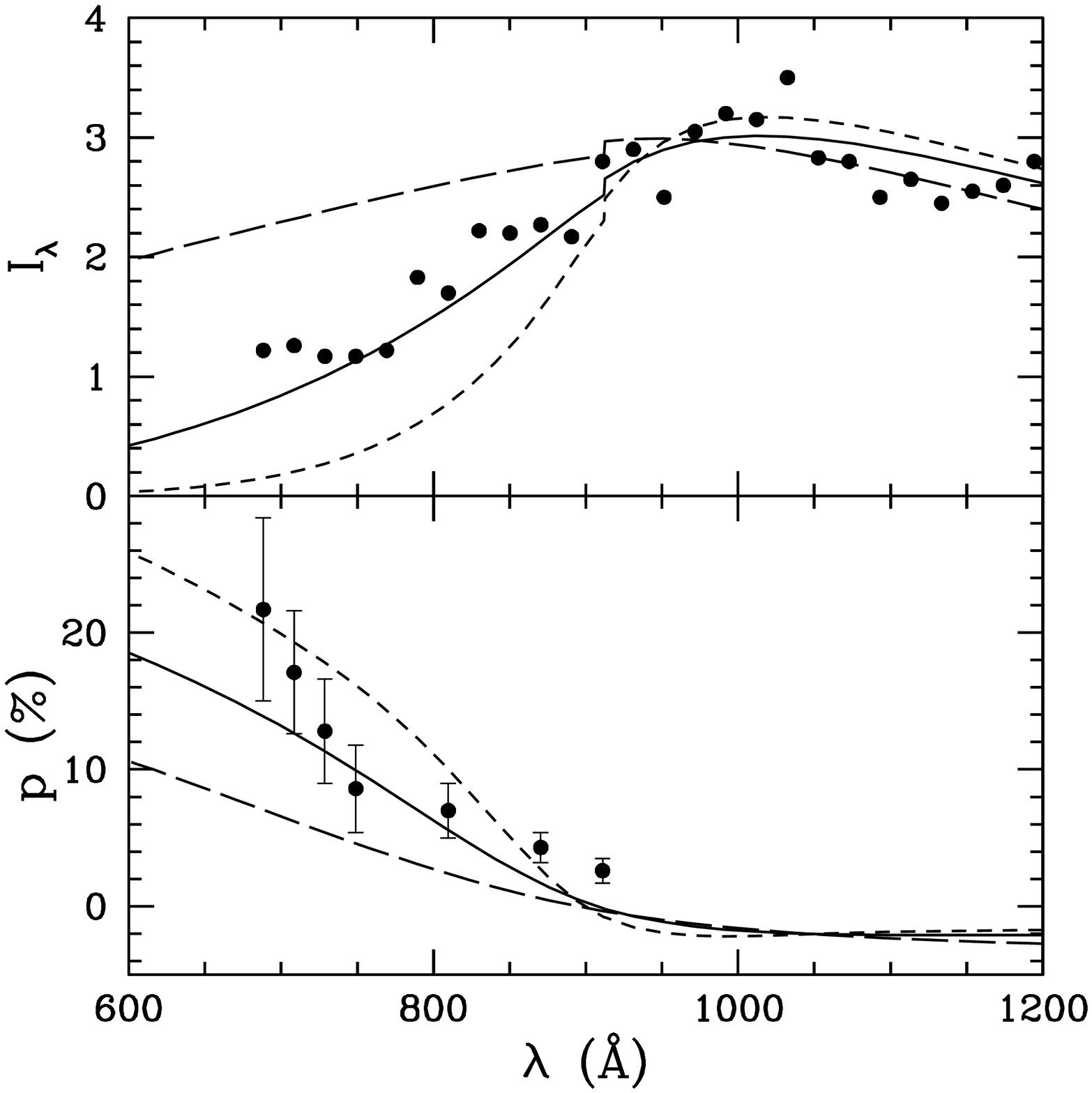}}
}
\figcaption{ Intensity, $I_\lambda$ (in arbitrary units), and polarization,
$p$, from a disk covered by an ionized ``skin'' of optical depth $\tT=1$, as 
viewed at inclination $\theta=77^{\rm o}$. The sign of $p$ is chosen so that 
$p>0$ corresponds to the case where the electric vector is parallel to the disk
normal. The dashed, solid, and long-dashed curves correspond to skin 
temperatures $kT=1$, 3, and 10 keV, respectively. Below the skin, we assume
unpolarized blackbody emission of temperature 3 eV with a sharp Lyman edge in 
absorption. The data for PG 1630+377 (Koratkar et al. 1995) show 
the detections of $p$ above the $2.5\sigma$ level.
\label{fig1}}
\bigskip

\smallskip
\centerline{\epsfxsize=8.4cm {\epsfbox{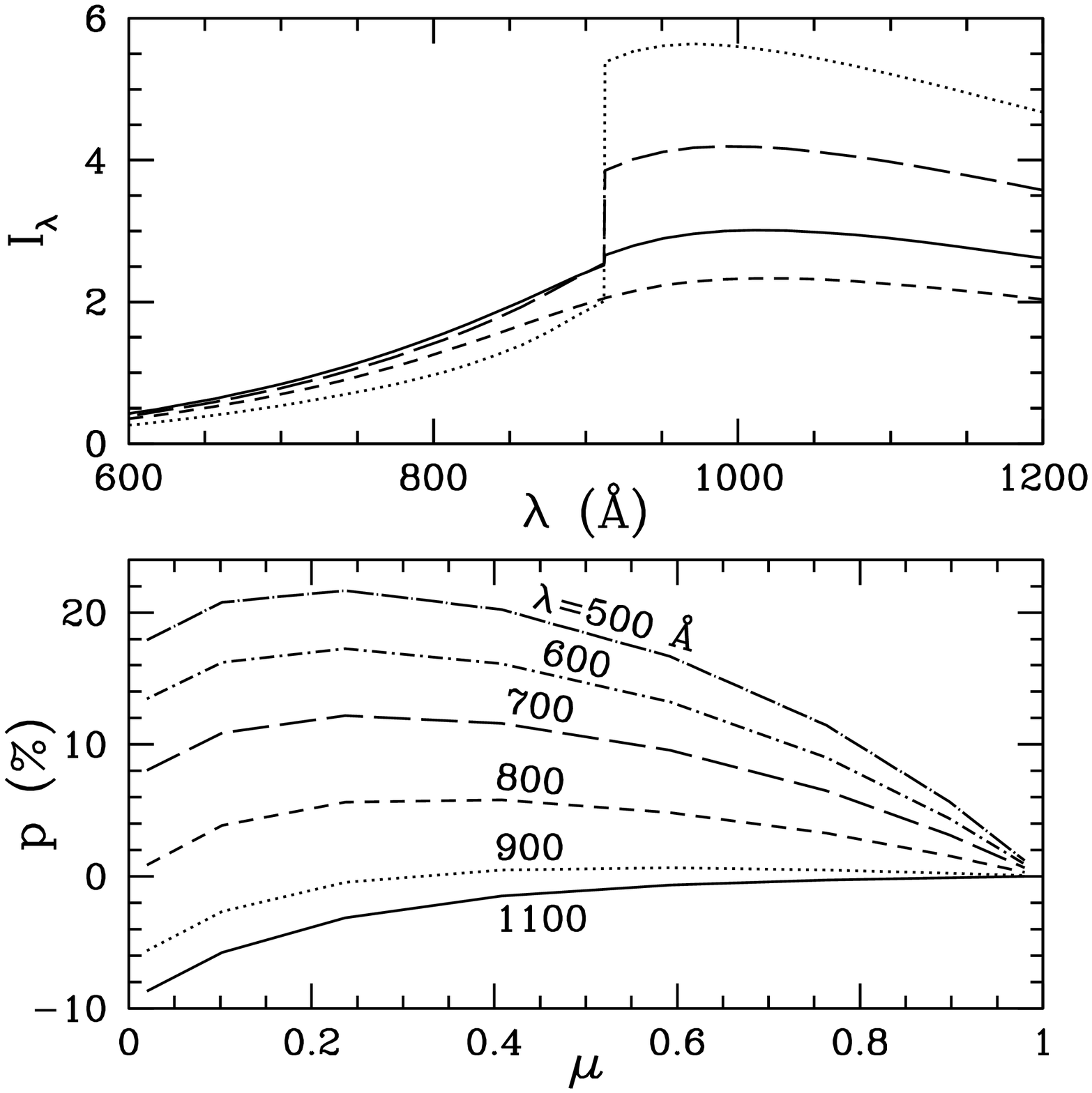}} }
\figcaption{ Skin model with $\tau_T=1$ and $kT=3$ keV. {\it Top:} 
Emerging intensity at different inclinations, $\mu=\cos\theta$. The dashed, 
solid, long-dashed, and dotted curves correspond to $\mu=0.047$, 0.23, 
0.5, and 0.95, respectively.
{\it Bottom:} Angular dependence of polarization at different wavelengths.
\label{fig2}}
\bigskip


\section{Scattering in a coronal outflow}

A corona above an accretion disk need not be static and formation of mildly 
relativistic outflows is very likely. Moreover, there are indications from 
X-ray observations for bulk motion in coronas of black hole sources 
(Beloborodov 1999a; Zdziarski, Lubi\'nski, \& Smith 1999). Here, we consider a 
toy model in which the outflow is replaced by a slab of outflowing plasma with 
a given column density, $N_c$. The underlying UV disk is assumed to emit 
isotropically. A characteristic velocity of the outflow is 
$\beta\equiv v/c\sim\beta_*=(4-\sqrt{7})/3\approx 0.45$, which corresponds to 
equilibrium with the disk radiation field (i.e., the bulk acceleration by the 
radiative pressure is balanced by the Compton drag at $\beta=\beta_*$,
Gurevich \& Rumyantsev 1965). The equilibrium must be established in an outflow
composed of $e^\pm$ pairs (Beloborodov 1999b). When viewed from the outflow 
comoving frame, the disk radiation is limb-brightened. This effect has a 
maximum at $\beta=\beta_*$ (B98).

A typical polarization of the scattered radiation is $\sim 10-20$\% depending
on the disk inclination, $\mu$ (B98). The observed weak polarization redward of
the Lyman edge, $p\simlt 2$\%, indicates that the contribution of the scattered 
radiation to the observed flux is small. In a slab geometry, it implies that 
the outflow is optically thin, $N_c<\sT^{-1}$. At the same time, blueward of 
the edge, the scattered radiation can dominate the observed flux, producing a 
steep rise in polarization at $\lambda<912$~{\AA}. If the primary radiation 
crucially decreases at $\lambda<912$ {\AA} then the flux here is mainly 
composed of scattered photons that have been blueshifted upon scattering due to
Doppler effect. In addition to the Doppler effect associated with bulk motion, 
there may be Doppler shifts due to thermal motions in the outflow. We therefore
consider outflows with a range of temperatures,~$T$.

\subsection{Calculations}

The equations of Thomson transfer of a polarized frequency-integrated radiation
in a cold outflow are given in B98. Here, we extend the problem: (1) We include
the dependence on frequency, $\nu$, and study the polarization along with the 
radiation spectrum. (2) We relax the assumption $kT\ll m_ec^2$ and calculate 
radiative transfer in outflows of high temperatures.
 
In a slab geometry, the polarized radiation is described by the two Stokes 
parameters, $I(\nu,\mu)$ and $Q(\nu,\mu)$.
The scattering is represented by the source functions, 
$S_I(\nu,\mu)$ and $S_Q(\nu,\mu)$. We hereafter use vector notation 
${\bf I}\equiv (I,Q)$ and ${\bf S}\equiv (S_I,S_Q)$. To calculate ${\bf S}$, we
transform ${\bf I}$ into the outflow comoving frame. 
Polarization, $p=Q/I$, is invariant with respect to Lorentz boosts along the 
normal, and ${\bf I}$ transforms as (see, e.g., Rybicki \& Lightman 1979)
$$
  {\bf I}^c(\nu_c,\mu_c)=D^3(\mu){\bf I}(\nu,\mu), 
$$
$$
  \mu_c=\frac{\mu-\beta}{1-\beta\mu}, \qquad 
  \nu_c=\gamma(1-\beta\mu)\nu\equiv D(\mu)\nu,
$$
where the index $c$ stands for the comoving frame and
$\gamma=(1-\beta^2)^{-1/2}$. In the comoving frame, we compute the source 
functions using the $2\times 2$ redistribution matrix, ${\bf R}$, 
derived for a hot static plasma (Nagirner \& Poutanen 1993; see also Poutanen 
\& Svensson 1996),
\begin{eqnarray}
\nonumber
  {\bf S}^c(\nu_c,\mu_c)=\int\int \left(\frac{\nu_c}{\nu_c^\prime}\right)^2
    {\bf R}(\nu_c,\mu_c,\nu_c^\prime,\mu_c^\prime)
    {\bf I}^c(\nu_c^\prime,\mu_c^\prime) \dd\nu_c^\prime\dd\mu_c^\prime,
\end{eqnarray}
and then transform ${\bf S}^c$ into the lab frame, 
${\bf S}(\nu,\mu)=D^{-3}(\mu){\bf S}^c(\nu_c,\mu_c)$. 

We calculate the observed radiation in the single-scattering approximation, 
i.e., ${\bf I}_{\rm obs}\approx {\bf I}+\tau_\mu({\bf S}-{\bf I})$, where
$\tau_\mu=N_c\sT(1-\beta\mu)/\mu$ is the slab optical depth at inclination 
$\mu$. This approximation is reasonable when (1) the outflow has a small 
optical depth ($\tau_\mu<1$) and (2) we are interested in the radiation near 
the Lyman edge. The high scattering orders will affect the results at energies
above a typical energy of once-scattered photons. 
We assume isotropic unpolarized primary radiation, ${\bf I}$,
(the primary depolarization in the disk is likely to be caused by Faraday 
rotation, e.g., Agol \& Blaes 1996).

\subsection{Results}

First, we consider the case of a sharp Lyman edge (Fig. 3). After passing 
through the outflow, a polarized tail appears in the spectrum blueward of the 
edge. One can see from Figure 3 how a high outflow temperature increases the 
flux at $\lambda<912$ {\AA}.
On the other hand, a high $T$ reduces the polarization of the upscattered 
radiation (see also Poutanen \& Vilhu 1993; Nagirner \& Poutanen 1994).  
This effect is due to mildly relativistic thermal motions that cause random 
aberration of light in the electron rest frame. The random character of the 
``thermal aberration'' leads to a reduction of the systematic polarization.

\medskip

\smallskip
\centerline{ \epsfxsize=8.4cm {\epsfbox{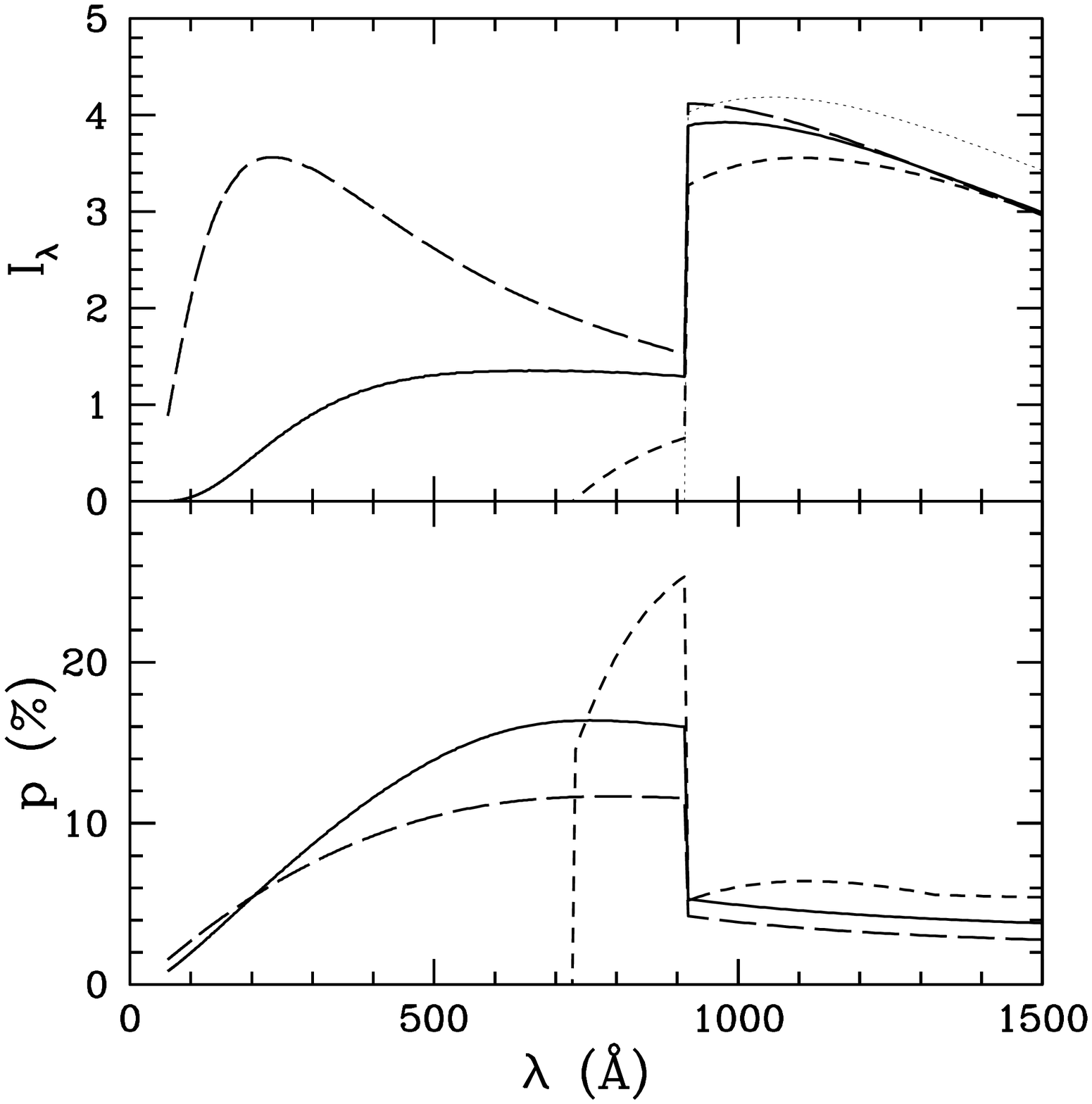}} }
\figcaption{ Disk radiation after passing through an outflow with 
$\beta=\beta_*=0.45$ and column density $N_c=0.2\sT^{-1}$, as viewed at 
$\mu=\beta$. $I_\lambda=I\nu^2/c$ is plotted in arbitrary units. The sign of 
$p$ is chosen so that $p>0$ corresponds to the parallel polarization.
The primary (unpolarized, isotropic) radiation of the disk is
assumed to have a blackbody spectrum, $kT_{\rm bb}=3$ eV, with a sharp Lyman 
edge ({\it dotted curve}).
The dashed, solid, and long-dashed curves represent the observed radiation 
(in the single-scattering approximation) for outflow temperatures $kT=0$, 50, 
and 100 keV, respectively.
\label{fig3}}
\bigskip

The amplitude of the tail is proportional to the outflow column density. The 
discontinuity in the observed spectrum comes from the edge in the primary 
radiation. The polarization is only due to scattered radiation which has
no discontinuity, hence $Q_{\rm obs}$ is smooth across the edge. This gives
a relation between the jumps in $p_{\rm obs}$ and $I_{\rm obs}$:
$p_{\rm obs}^+/p_{\rm obs}^-=I_{\rm obs}^-/I_{\rm obs}^+$,
where $+/-$ correspond to $\lambda$ just above/below 912 {\AA}.

In order to emulate the observed spectrum of PG 1630+377, we use a toy 
primary intensity $I_\ep\propto\ep^2\exp[-(\ep/\ep_0)^6]$ and 
choose $\ep_0=13$ eV. This spectrum emulates a smeared-out Lyman edge 
(shown by the dotted curve in Fig. 4). The original smearing-out may be 
the result of relativistic effects near the black hole. 
The emerging radiation after passing through the outflow is shown in Figure 4 
for $N_c=0.23\sT^{-1}$ at inclination $\theta=60^{\rm o}$. 
A reasonably good fit to the PG 1630+377 data is obtained for $kT\approx 
50-100$ keV. 

\medskip

\smallskip
\centerline{
\epsfxsize=8.4cm {\epsfbox{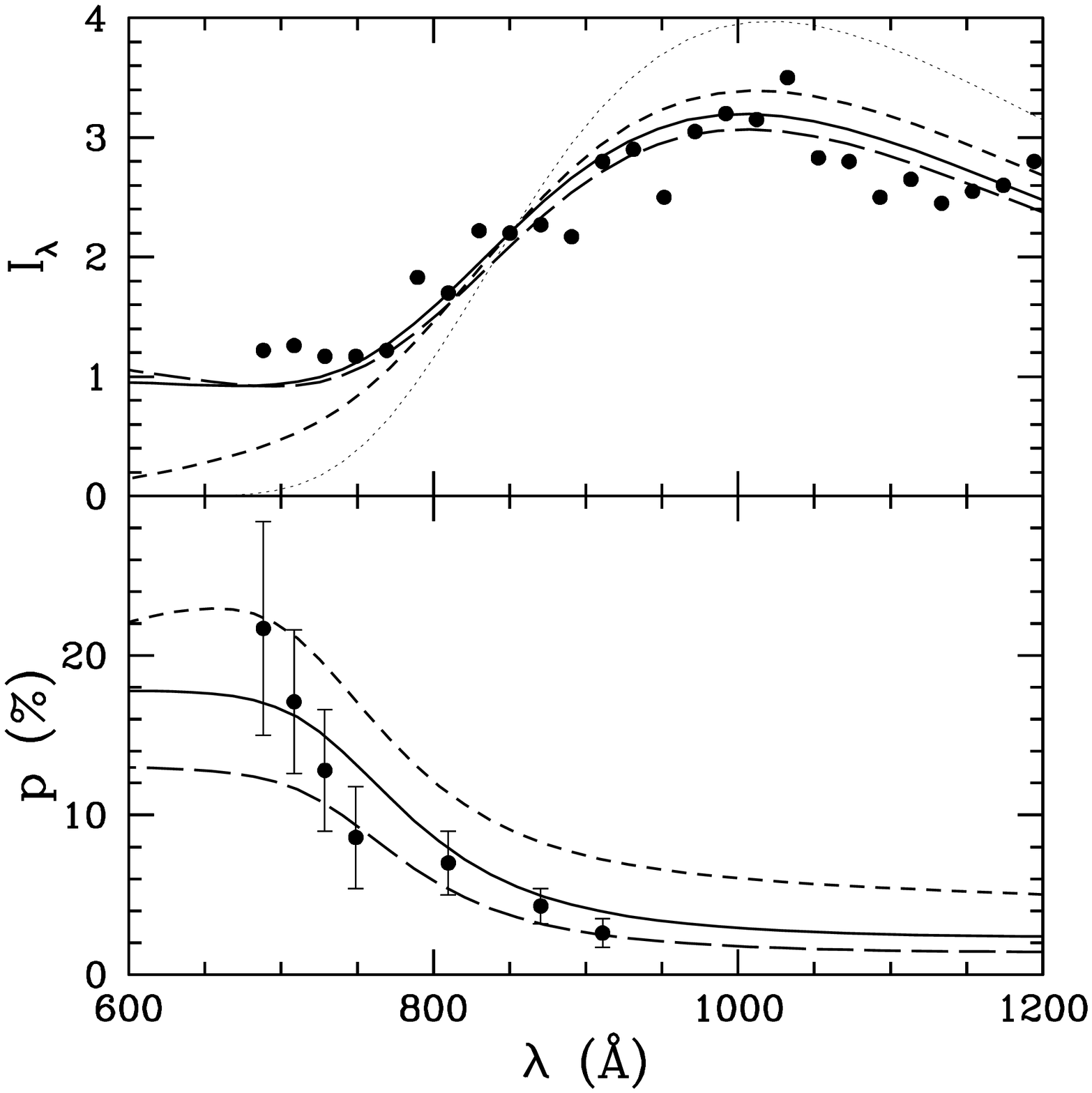}}
}
\figcaption{ Outflow model with $\beta=\beta_*=0.45$ and column density 
$N_c=0.23\sT^{-1}$, as viewed at inclination $\theta=60^{\rm o}$, against
the data on PG 1630+377.
The assumed original disk emission is shown by the dotted curve.
The dashed, solid, and long-dashed curves correspond to 
outflow temperatures $kT=3$, 50, and 100 keV, respectively.
\label{fig4}}
\bigskip

A special feature of the outflow model is that the maximum $p$ is 
predicted at $\mu=\beta$ which corresponds to $\mu_c=0$, albeit the 
polarization is also high for a wide range of inclinations (see B98).
More importantly, the parallel orientation is predicted at all wavelengths,
in contrast to our first (static) model. Note also that the outflow model 
provides a steeper rise in $p$ and predicts a hardening in the spectrum at 
$\sim 750$ {\AA} similar to that present in the data.


\section{Summary}

We have found that Comptonization of UV radiation with a Lyman edge in 
absorption can produce very steep polarization rises blueward of the edge. 
Redward, the predicted polarization is low and weakly dependent on wavelength.

We model the Comptonizing plasma as a hot slab with three parameters: 
(1) the optical depth, $\tT$, (2) the temperature, $T$, and (3) the bulk 
velocity, $\beta$. Our first (static) model emulates an ionized skin of an 
accretion disk.  It has $\tT\sim 1$, $kT\sim 3$ keV, and $\beta=0$. 
The skin produces a steep rise in parallel polarization blueward of the 
Lyman edge. The predicted polarization, however, changes its sign across the 
edge, which is not favored by the observations. Another potential difficulty 
is the strong Lyman edge predicted at modest inclinations.

In our second (outflow) model, the hot plasma is optically thin, $\tT<1$, 
and its bulk velocity, $\beta\sim 0.5$, is in equilibrium with the disk 
radiation field (B98). A steep rise in  polarization is then generated for a 
wide range of parameters. The predicted polarization is parallel to the disk 
normal at all wavelengths. Note that the redward polarization may be further 
reduced if an additional unpolarized source contributes at large wavelengths 
(e.g., the disk emission from larger radii, which is not scattered in the 
central outflow). 

Both models predict polarization rises only in those objects where there is 
a Lyman edge or a dramatic spectral break in the original disk spectrum.
A large polarization, $p\simgt 20$\%, is then predicted at favorable 
inclinations. The skin model has maximum $p$ at $\cos\theta\approx 0.25$, 
and the outflow model -- at $\cos\theta=\beta$.

The outflow model appears to be more promising. 
Note that the slab geometry is a rough approximation for an 
outflow. A real outflow is at least 2D (axisymmetric) or
it consists of  localized ejecta from the disk (Beloborodov 1999a).
A full 2D model should include self-consistently 
the effects of relativistic disk rotation and gravitational redshift, which
are important near a Kerr black hole.

\acknowledgments

This work was supported by the Swedish Natural Science Research Council
and RFFI grant 97-02-16975.


\begin{references}

\reference{}
Agol, E., \& Blaes, O. 1996, MNRAS, 282, 965

\reference{}
Antonucci, R.R.J. 1992, in Testing the AGN paradigm, ed. S. Holt,
S. Neff, C.M. Urry (New York), 486

\reference{}
Beloborodov, A. M. 1998, ApJ, 496, L105 (B98)

\reference{}
Beloborodov, A. M. 1999a, ApJ, 510, L123

\reference{}
Beloborodov, A. M. 1999b, MNRAS, in press

\reference{}
Blaes, O., \& Agol, E. 1996, ApJ, 469, L41

\reference{}
Gurevich, L. E., \& Rumyantsev, A. A. 1965, Sov. Physics -- JETP, 20, 1233

\reference{}
Haardt, F., Maraschi, L., \&  Ghisellini, G. 1994, ApJ, 432, L95

\reference{}
Haardt, F., \& Matt, G. 1993, MNRAS, 261, 346

\reference{}
Hsu, C.-M., \& Blaes, O. 1998, ApJ, 506, 658

\reference{}
Impey, C. D., Malkan, M. A., Webb, W., \& Petry, C. E. 1995, ApJ, 440, 80

\reference{}
Koratkar, A., Antonucci, R. R. J., Goodrich, R. W., Bushouse, H., \& 
Kinney, A. L. 1995, ApJ, 450, 501

\reference{}
Koratkar, A., \& Blaes, O. 1999, PASP, 111, 1

\reference{}
Nagirner, D. I., \& Poutanen, J. 1993, A\&A, 275, 325

\reference{}
Nagirner, D. I., \& Poutanen, J. 1994, Astrophys. Space Phys. Reviews, 9, 1

\reference{}
Poutanen, J., \& Svensson, R. 1996, ApJ, 470, 249

\reference{}
Poutanen, J., \& Vilhu, O. 1993, A\&A, 275, 337

\reference{}
Rybicki, G. B., \& Lightman, A. P. 1979, Radiative Processes in Astrophysics,
New York: Wiley

\reference{}
Shields, G. A., Wobus, L., \& Husfeld, D. 1998, ApJ, 496, 743

\reference{}
Sunyaev, R. A., \& Titarchuk, L. G. 1985, A\&A, 143, 374


\reference{}
Zdziarski, A. A., Lubi\'nski, P., \& Smith, D. A. 1999, MNRAS, 303, L11

\reference{}
\.Zycki, P. T., Krolik, J. H., Zdziarski, A. A., \& Kallman, T. R. 1994,
ApJ, 437, 597



\end{references}
\end{document}